\documentclass[preprint, 12pt, aps, prl, superscriptaddress]{revtex4-2}
\usepackage{graphicx}
\usepackage{booktabs,array}
\usepackage{braket}
\usepackage{amsmath}
\usepackage{subcaption}
\usepackage{caption}
\usepackage{color,soul}
\usepackage[dvipsnames]{xcolor}

\usepackage{hyperref}
\usepackage{subfiles}

\newcommand{\colorsout}[1]{\bgroup\markoverwith{\textcolor{#1}{\rule[0.5ex]{2pt}{0.4pt}}}\ULon}

\newcommand{\expSTMESR}{~\cite{Baumann_Paul_science_2015,
Natterer_Yang_nature_2017,Choi_Paul_natnano_2017,Willke_Paul_sciadv_2018,Yang_Bae_prl_2017,
Willke_Bae_science_2018,Y_Bae_advanced_science_2018,Willke_Singha_nanolett_2019,
Willke_Yang_natphys_2019,Yang_Paul_prl_2019,yang_coherent_2019,Seifert_Kovarik_pr_2020,
Seifert_Kovarik_eabc_2020,
Weerdenburg_Steinbrecher_2020,Steinbrecher_Weerdenburg_2020,Jinkyung_Hyperfine_2022,hong_2024_acsnano}}

\usepackage{xr}  % Add this line to use the xr package
\newcommand{\ourtitle}{Spin-State Engineering of Single Titanium Adsorbates on Ultrathin Magnesium Oxide}

\begin{document}

% Journals: Nat. Comm. -> ACS Nano

\author{Soo-hyon Phark}
\email{phark@qns.science}
\affiliation{Center for Quantum Nanoscience, Institute for Basic Science (IBS), Seoul 03760, Republic of Korea}
\affiliation{Ewha Womans University, Seoul 03760, Republic of Korea}

\author{Hong Thi Bui}
\affiliation{Center for Quantum Nanoscience, Institute for Basic Science (IBS), Seoul 03760, Republic of Korea}
\affiliation{Department of Physics, Ewha Womans University, Seoul 03760, Republic of Korea}

\author{We-hyo Seo}
\affiliation{Center for Quantum Nanoscience, Institute for Basic Science (IBS), Seoul 03760, Republic of Korea}
\affiliation{Ewha Womans University, Seoul 03760, Republic of Korea}

\author{Yaowu Liu}
\affiliation{Center for Quantum Nanoscience, Institute for Basic Science (IBS), Seoul 03760, Republic of Korea}
\affiliation{Ewha Womans University, Seoul 03760, Republic of Korea}

\author{Valeria Sheina}
\affiliation{Center for Quantum Nanoscience, Institute for Basic Science (IBS), Seoul 03760, Republic of Korea}
\affiliation{Ewha Womans University, Seoul 03760, Republic of Korea}

\author{Curie Lee}
\affiliation{Center for Quantum Nanoscience, Institute for Basic Science (IBS), Seoul 03760, Republic of Korea}
\affiliation{Department of Physics, Ewha Womans University, Seoul 03760, Republic of Korea}

\author{Christoph Wolf}
\email{wolf.christoph@qns.science}
\affiliation{Center for Quantum Nanoscience, Institute for Basic Science (IBS), Seoul 03760, Republic of Korea}
\affiliation{Ewha Womans University, Seoul 03760, Republic of Korea}

\author{Andreas J. Heinrich}
\affiliation{Center for Quantum Nanoscience, Institute for Basic Science (IBS), Seoul 03760, Republic of Korea}
\affiliation{Department of Physics, Ewha Womans University, Seoul 03760, Republic of Korea}

\author{Roberto Robles}
\affiliation{Centro de F{\'{i}}sica de Materiales CFM/MPC (CSIC-UPV/EHU), 20018 Donostia-San Sebasti\'an, Spain}

\author{Nicol{\'a}s Lorente}
\email{nicolas.lorente@ehu.eus}
\affiliation{Centro de F{\'{i}}sica de Materiales CFM/MPC (CSIC-UPV/EHU), 20018 Donostia-San Sebasti\'an, Spain}
\affiliation{Donostia International Physics Center (DIPC), 20018 Donostia-San Sebasti\'an, Spain}

\title{\ourtitle}

\begin{abstract}
Single atomic adsorbates on ultrathin insulating films provide a promising route toward bottom-up quantum architectures based on atomically identical yet individually addressable spin qubits on solid surfaces. A key challenge in engineering quantum-coherent spin nanostructures lies in understanding and controlling the spin state of individual adsorbates. In this work, we investigate single titanium (Ti) atoms adsorbed on MgO/Ag(100) surfaces using a combined scanning tunneling microscopy and electron spin resonance. Our measurements reveal two distinct spin states, $S = 1/2$ and $S = 1$, depending on the local adsorption site and the thickness of the MgO film. Density functional theory calculations suggest a Ti$^+$ configuration for the Ti adsorbates with approximately 3 electrons in the 4$s$ and 3$d$ valence shells. Using a multi-orbital atomic multiplet calculations the site dependence of the spin can be rationalized as a charge redistribution between spin-polarizing and depolarizing orbitals. These findings underscore the potential of surface-supported single atoms as spin qubits with tunable spin and charge states, enabling atom-by-atom control in the realization of a versatile quantum platform on surfaces.
\end{abstract}

\maketitle
\clearpage

\section{Introduction}
\noindent
Positioning individual atoms using the tip of a scanning tunneling microscope (STM) has paved the way for the bottom-up fabrication of tailored nanostructures on surfaces with atomic precision.\cite{Eigler_Schweize_nature_1990} Subsequent achievements highlighted the potential of single atoms on surfaces as basic building blocks for constructing nanoscale devices with diverse charge and spin functionalities using the atomic precision of the STM.\cite{Baumann_Paul_science_2015,Natterer_Yang_nature_2017} The implementation of electron spin resonance (ESR) in STM experiments has pushed the energy resolution to an extraordinary limit ($\sim$10 neV)\expSTMESR, and has progressed towards pulsed manipulation of individual spin states since the first unequivocal continuous-wave ESR spectra were obtained from iron (Fe) single adsorbates on two atomic layers (2~ML) of magnesium oxide (MgO) on a silver substrate.\cite{Baumann_Paul_science_2015, Yang2019a}

Atoms or molecules hosting an electronic spin $S = 1/2$, when coupled to a magnetic field, serve as natural two-level systems and are promising candidates for building a solid-surface quantum platform of identical but individually addressable qubits.\cite{heinrich_nnano_2021,wolfphark2024,DJChoi2025_RSC,phark2025_roadmap} In this context, ESR-STM studies on a variety of $S = 1/2$ systems, including titanium (Ti),\cite{yang_PRL_2017} iron(II) phthalocyanine (FePc),\cite{zhang_electron_2022,willke_FePc_2021} alkali dimers and pairs on MgO,\cite{Kovarik_2022}, molecular radicals,\cite{Kovarik2024SpinMolecule,czap2025magneticresonanceimagingsingle} and bis(phthalocyaninato)terbium(III) on sodium chloride (NaCl),\cite{Kawaguchi_2022} have demonstrated the suitability of single-spin adsorbates for quantum-coherent science at the atomic scale. In particular, the use of Ti atoms on MgO has enabled the realization of advanced ESR control protocols, including universal single-qubit gates, a two-qubit gate,\cite{Wang2023_vector, wang2023} and the free coherent evolution of entangled spins.\cite{Veldman}

Despite these advances, the practical implementation of atomic spin qubits in quantum technologies remains challenging. This is due, in large part, to the complex interplay between adsorbates and the substrate, which can induce deviations from free-atom configurations, both in charge and spin states, and may result in reduced coherence times. A detailed understanding of the physical and chemical environment of single atoms on surfaces is therefore essential for the controlled realization of scalable quantum-coherent nanostructures.

The charge and spin states of individual adsorbates can differ significantly from their isolated counterparts, due to various mechanisms such as charge transfer, ligand field effects, or chemical reactions during deposition. For example, Ti atoms—expected to have $S = 1$ in the gas phase—have been found to exhibit $S = 1/2$ on MgO/Ag(100), previously attributed to hydrogen capture from residual gas.\cite{yang_PRL_2017, Willke_Bae_science_2018, yang_coherent_2019, Steinbrecher_2021} Similarly, FePc has been observed to change its spin state from $S = 1$ to $S = 1/2$ due to electron transfer from the Ag substrate.\cite{willke_FePc_2021, zhang_electron_2022} A systematic understanding of how the substrate modifies the electronic configuration of adsorbates is thus a prerequisite for designing robust atomic-scale quantum systems.

In this work, we combine scanning tunneling spectroscopy (STS), ESR, density functional theory (DFT) and multiplet calculations to investigate the spin and charge states of Ti atoms adsorbed on MgO films of 2 and 3 monolayers (MgO$^{(2)}$ and MgO$^{(3)}$ respectively) formed on Ag(100). We observe that the spin state of the adsorbate depends sensitively on both the local adsorption geometry and the MgO thickness. Specifically, Ti exhibits a spin $S = 1/2$ when adsorbed on both oxygen-atop (O-atop) and O-O bridge sites of MgO$^{(2)}$, and on bridge sites of MgO$^{(3)}$. In contrast, on O-atop sites of MgO$^{(3)}$, we find a spin of $S > 1/2$, identified via remote-spin ESR, STS and multiplet calculations. Atom manipulation confirms robust and reversible switching between these distinct spin states and adsorption geometries. Our DFT calculations support a consistent Ti$^+$ oxidation state primarily governed by the lowering of the work function by the MgO layer, the actual Ti spin can be rationalized as a charge redistribution between spin polarizing and depolarizing orbitals that we model by 4$s^2$3$d^1$ and 4$s^1$3$d^2$ configurations in our magnetic multiplet calculations. Our findings indicate that the interaction on the adsorption site, rather than chemical doping or hydrogenation, is the principal mechanism for spin variability, which provides new venues for the quantum engineering of surface-supported spin qubits.
\section{Experiment}

\subsection{STM images and STS spectra}
\noindent
The sample was prepared by depositing Ti and Fe atoms with a submonolayer coverage ($< 1\%$) onto defect-free MgO films grown on a clean Ag(100) substrate (see Methods and Supporting Information S1.A). Figure~\ref{fig1}A presents an STM topography of the sample surface, which includes regions of MgO$^{(2)}$, MgO$^{(3)}$, and bare Ag. Well-isolated atomic adsorbates are observed throughout all regions.

High-resolution STM images on MgO$^{(2)}$ (Fig.~\ref{fig1}B) reveal two distinct Ti adsorption geometries: one atop surface oxygen atoms (Ti$_\mathrm{O}^{(2)}$), and one bridging two adjacent oxygen atoms (Ti$_\mathrm{B}^{(2)}$), consistent with earlier reports~\cite{Y_Bae_advanced_science_2018,Kim_2021}. Similarly, on MgO$^{(3)}$ (Fig.~\ref{fig1}C), two Ti species are also identified. The bridge-site configuration (Ti$_\mathrm{B}^{(3)}$) appears with considerable abundance and shows STS characteristics closely resembling those of Ti$_\mathrm{B}^{(2)}$. In contrast, the oxygen-atop configuration (Ti$_\mathrm{O}^{(3)}$) displays a much larger apparent height in STM images and distinct spectroscopic features (Fig.~\ref{fig1}E), which are discussed in detail below. A third, minor O-atop variant, similar to Ti$_\mathrm{O}^{(2)}$, was also occasionally observed on MgO$^{(3)}$ but accounted for less than 0.1\% of the total Ti adsorbate population. This minority species is not considered further (see Supporting Information Fig. S2).

To probe the electronic properties of the four dominant Ti configurations identified in STM (Figs.~\ref{fig1}B and C), we performed STS measurements (Figs.~\ref{fig1}D and E). The spectra acquired for Ti adsorbates on MgO$^{(2)}$ exhibit features in agreement with previous studies~\cite{Yang_Bae_prl_2017,Seifert_Kovarik_eabc_2020,Veldman}. Likewise, Ti$_\mathrm{B}^{(3)}$ exhibits a nearly identical spectral profile to Ti$_\mathrm{B}^{(2)}$, characterized by a smooth, monotonic conductance curve and a step-like feature at zero bias (marked by a cyan triangle), attributed to spin polarization of the STM tip~\cite{Kim_2021,Hwang_2022}. In contrast, Ti$_\mathrm{O}^{(3)}$ exhibits a pronounced inelastic tunneling (IET) signature at $\pm 18$~meV (orange arrows in the inset), symmetrically positioned about $V_\mathrm{b} = 0$. This feature indicates a zero-field splitting and the presence of a sizable magnetic anisotropy energy (MAE), consistent with a spin state $S > 1/2$. DFT and multiplet calculations for Ti$_\mathrm{O}^{(3)}$ confirm that this state corresponds to a magnetic ground state with spin $S = 1$ (see the theory section below).

\subsection{ESR spectra}
\noindent
ESR measurements of the Ti adsorbates provided direct insight into their spin states (Fig.~\ref{fig2}A). Three species—Ti$_\mathrm{O}^{(2)}$, Ti$_\mathrm{B}^{(2)}$, and Ti$_\mathrm{B}^{(3)}$—exhibit ESR resonances (red, blue, and cyan traces, respectively) within the frequency range expected for Zeeman splitting under an applied magnetic field of $\sim 0.6$~T and an effective $g$-factor of $\sim 2$, consistent with a spin angular momentum of $S = 1/2$~\cite{Yang_Bae_prl_2017,Kim_2021}. The assignment of $S = 1/2$ for Ti$_\mathrm{B}^{(3)}$ is further supported by the close resemblance between its \textit{d}I/\textit{d}V spectrum (Fig.~\ref{fig1}D, blue) and that of Ti$_\mathrm{B}^{(2)}$ (Fig.~\ref{fig1}E, cyan). 

In contrast, Ti$_\mathrm{O}^{(3)}$ does not display any ESR resonance within the same frequency window (orange trace in Fig.~\ref{fig2}A), indicating a fundamentally different spin configuration. Coupled with the IET-like steps observed in its \textit{d}I/\textit{d}V spectrum at $\pm 18$~meV (Fig.~\ref{fig1}E), this suggests a spin state with $S > 1/2$ and significant magnetic anisotropy, warranting further investigation.

To probe the magnetic anisotropy of Ti$_\mathrm{O}^{(3)}$, we analyzed its interaction with a neighboring spin. In this approach, the target spin (Ti$_\mathrm{O}^{(3)}$) couples to a nearby probe spin with $S = 1/2$ (Ti$_\mathrm{B}^{(3)}$), and the anisotropy of the target is inferred from angular variations in the ESR spectrum of the probe~\cite{Phark_2023,Reale2024}. The Ti$_\mathrm{B}^{(3)}$–Ti$_\mathrm{O}^{(3)}$ pair was assembled by lateral manipulation using the STM tip (`dragging'; see Fig.~\ref{fig3}C), with a separation of 0.72~nm as denoted in Fig.~\ref{fig2}B.

The ESR spectrum acquired on the Ti$_\mathrm{B}^{(3)}$ in this pair (Fig.~\ref{fig2}C) exhibits a clear peak splitting $\Delta f$ arising from spin–spin interaction, modeled as $J \vec{S}_\mathrm{B} \cdot \vec{S}_\mathrm{O}$. The angular dependence of $\Delta f$ with respect to the orientation of the external magnetic field $\theta_{\rm ext}$ (Fig.~\ref{fig3}D) follows a $\cos \theta_{\rm ext}$ trend~\cite{Phark_2023}, consistent with uniaxial out-of-plane anisotropy of the Ti$_\mathrm{O}^{(3)}$ spin. Given that the Ti$_\mathrm{B}^{(3)}$ spin aligns with the external field direction, the sinusoidal modulation of $\Delta f(\theta_{\rm ext})$ provides strong evidence for a spin angular momentum $S_\mathrm{O} > 1/2$ in Ti$_\mathrm{O}^{(3)}$. The observation of two sinusoidal branches further suggests bistability in the spin orientation of Ti$_\mathrm{O}^{(3)}$, with preferred alignment along two collinear directions, which in turn defines the pair of resonance frequencies observed in the ESR spectrum (see up and down arrows in Fig.~\ref{fig3}C).

\subsection{Control of adsorption sites}
\noindent
When adsorbed on MgO$^{(2)}$, Ti atoms exhibit relatively short spin lifetimes and coherence times, most likely limited by substrate-induced spin scattering in the absence of tunneling electrons.~\cite{Paul_Yang_natphys_2017} Consequently, Ti atoms adsorbed on thicker insulating layers such as MgO$^{(3)}$ may exhibit significantly enhanced coherence, owing to the reduced electronic coupling to the underlying Ag substrate, which suppresses backscattering of free electrons~\cite{Paul_Yang_natphys_2017}. In this context, achieving precise control over Ti atom positioning on MgO$^{(3)}$ is essential for constructing artificial spin structures with optimized quantum coherence. To this end, we employ two complementary atom manipulation techniques to reposition Ti atoms on MgO$^{(3)}$, as illustrated in Fig.~\ref{fig3} (see Supporting Information S2 for technical details).

The first method consists of picking up a Ti$_\mathrm{O}^{(3)}$ atom with the STM tip and dropping it off at a new location, as shown schematically in Fig.~\ref{fig3}A. This approach is suitable for long-range repositioning, typically resulting in a Ti$_\mathrm{O}^{(3)}$ species within a MgO$^{(3)}$ domain. Interestingly, when the atom is dropped onto a MgO$^{(2)}$ region, the resulting species is consistently ESR-active, corresponding to either Ti$_\mathrm{O}^{(2)}$ or Ti$_\mathrm{B}^{(2)}$, as confirmed by both STM contrast and spectroscopic features (see Fig.~\ref{fig3}B for Ti$_\mathrm{O}^{(2)}$). Given the well-established spin $S = 1/2$ character of Ti adsorbates on MgO$^{(2)}$, this observation supports the assignment of a spin $S = 1$ to Ti$_\mathrm{O}^{(3)}$, in agreement with the configuration expected for a neutral Ti atom in vacuum.

The second method enables short-range repositioning by inducing lateral hopping of the Ti atom between neighboring lattice sites using voltage pulses applied to the STM junction, as depicted in Fig.~\ref{fig3}C (`dragging'). By fine-tuning the lateral tip displacement $x_{\rm hop}$, vertical approach $z_{\rm hop}$, and pulse amplitude $V_{\rm hop}$, we achieve controlled atomic movement over half the O–O lattice spacing on the MgO surface (see sequential STM images in Fig.~\ref{fig3}D). This manipulation protocol enables precise construction of multi-atom assemblies, including the Ti$_\mathrm{B}^{(3)}$–Ti$_\mathrm{O}^{(3)}$ pair discussed in Fig.~\ref{fig2}B. Notably, the hopping process is reversible, allowing the Ti atom to switch between two distinct adsorption sites. As demonstrated in Figs.~\ref{fig1}E and~\ref{fig2}A, this transition induces a corresponding change in the atom’s spin, which is either $S = 1$ or $S = 1/2$, depending on whether the atom occupies an O-atop or bridge site. This strongly suggests that the spin state of Ti is governed by its local bonding environment and can be reversibly switched via site-selective manipulation.

\section{Theory}
\noindent
Such a reversible change of spin states is unlikely to be the result of selective adsorption or desorption of atomic hydrogen (H). In general, H-adsorption has been suggested to occur in experimental conditions under ultrahigh vacuum and low temperatures for Ti adsorbates.~\cite{Yang_Bae_prl_2017} Indeed, our DFT calculations show that Ti can readily form stable hydride species when adsorbed on the MgO/Ag(100) surface, see the Supplementary Information. Among the considered configurations, TiH$_2$ emerges as the thermodynamically favored species under hydrogen-rich conditions, as determined by comparing Gibbs free energies referenced to gas-phase H$_2$. The formation of TiH$_3$ is found to be unstable, typically leading to spontaneous H desorption. 

 TiH has a reduced thermodynamic phase space and can only be formed for a small range of hydrogen chemical potentials. Under low concentration of hydrogen, bare Ti will be stabilized depending on the source of hydrogen that fixes the hydrogen chemical potential. This is further supported by the experimental observation that Ti adatoms can slowly degrade over time, ultimately becoming spectroscopically dark not exhibiting any ESR features. We assign this latter behavior to a degradation process related to slow hydrogen adsorption that happens on the scale of months.

Under ultra-low temperature and ultrahigh vacuum experimental conditions, Ti atoms are likely found on the surface.  A comparison of the theoretically obtained magnetic structures of the TiH$_x$ species indicates that the bare Ti adsorbate is compatible with the experimental data, while TiH does not reproduce the observed behavior (see the Supplementary Information for details). 

Our calculations give us access to the adsorption energies, which can be tentatively compared to the relative abundance of adsorbate on the atop and bridge sites. In our experiments, the prevalence of the bridge site on two monolayers is changed to the atop site on three monolayers. This behavior is only reproduced by the chemisorption energy of the Ti adsorbate (Supplementary Information).

In the following, we analyze in detail the electronic structure of Ti atoms on MgO$^{(2)}$ and MgO$^{(3)}$.

\subsection{Ti on MgO/Ag(100): Electronic Configuration}
\noindent
Our DFT calculations reveal distinct behavior depending on the adsorption site of Ti (O-atop vs.\ O-bridge), whereas the influence of MgO thickness (MgO$^{(2)}$ vs.\ MgO$^{(3)}$) is comparatively minor. Based on the experimental evidence, the most probable surface species is a single Ti atom in a positively charged Ti$^+$ configuration, resulting from charge transfer to the Ag substrate, facilitated by the reduced work function due to the MgO overlayer.

In all cases, charge analysis indicates that Ti adsorbed on MgO/Ag(100) retains approximately three valence electrons (see Table~\ref{Table_charges}). While the non-integer values arise from the intrinsic ambiguity in decomposing the DFT total charge density into orbital contributions, the net charge consistently supports a Ti$^+$ oxidation state. 

\begin{table*}[h]
\centering
\begin{tabular}{ c| c| c| c| c }
& Total Charge & 3$d$\ (up; down) & 4$s$\ (up; down) & $M_\mathrm{Ti}$ ($\mu_{\rm B}$) \\ 
 \hline
 Ti$_{\rm B}^{(2)}$ & 3.02  & 2.31 (1.51; 0.80) & 0.71 (0.28; 0.43) & 0.87 \\  \hline
 Ti$_{\rm O}^{(2)}$ & 3.24  & 2.42 (2.24; 0.18) & 0.82 (0.53; 0.29) & 2.31 \\  \hline
 Ti$_{\rm B}^{(3)}$ & 3.03  & 2.31 (1.89; 0.42) & 0.72 (0.25; 0.47) & 1.25 \\ \hline 
 Ti$_{\rm O}^{(3)}$ & 3.16  & 2.45 (2.24; 0.21) & 0.71 (0.49; 0.22) & 2.32 \\
\end{tabular}
\caption{\textbf{Electronic structure of Ti on MgO/Ag(100):} The table reports the total ($s$+$d$) valence charge, the spin-resolved occupations of the 3$d$ and 4$s$ orbitals, and the total magnetic moment of Ti adsorbed on different MgO sites and thicknesses. While the total occupations remain similar across configurations, the degree of spin polarization (and hence the effective spin state) varies markedly with adsorption site. See text and Supplementary Information for further discussion.}
\label{Table_charges}
\end{table*}

As shown in Table~\ref{Table_charges}, the total 3$d$ occupation lies consistently in the range of $d^{2.3}$–$d^{2.45}$, with the 4$s$ orbital slightly less than singly occupied. Despite the similar total electron count, the magnetic moments differ significantly: O-atop configurations yield magnetic moments of $\sim 2.3~\mu_{\rm B}$, while bridge-site configurations exhibit reduced moments as low as $0.87~\mu_{\rm B}$ on MgO$^{(2)}$. This disparity arises not from charge transfer but from spin depolarization within the occupied states.

Figure~\ref{PDOS_Ti} displays the projected density of states (PDOS) for Ti adsorbed on MgO/Ag(100). O-atop adsorption (Figs.~\ref{PDOS_Ti}(A) and (B)) shows two occupied $d$-orbitals around -1 eV, which yield a magnetic moment close to $2~\mu_{\rm B}$ (Table~\ref{Table_charges}) for both MgO thicknesses. The 4$s$ orbital plays only a minor role; instead, two $d$-orbitals in the majority-spin channel are occupied, resulting in an effective spin $S = 1$. This finding aligns well with the experimental results for MgO$^{(3)}$, but differs for MgO$^{(2)}$, where ESR indicates an $S = 1/2$ configuration.

In  bridge-site geometries (Figs.~\ref{PDOS_Ti}(C) and (D)), the $d$-orbitals appear at approximately $-1$~eV below the Fermi level. Depolarizing contributions appear near the Fermi level. These results are consistent with a $d^{2.3}$ configuration of Table~\ref{Table_charges} with a magnetic moment close to $S=1/2$.

To further elucidate the observed spin states and particularly the contrasting behavior of Ti on O-atop sites of MgO$^{(2)}$ versus MgO$^{(3)}$, we performed magnetic multiplet calculations assuming three valence electrons in the $s$ and $d$ shells. In this model, the 4$s$ orbital acts as a depolarizing channel, mimicking the enhanced hybridization and spin screening expected for bridge-site adsorption. This approach enables us to capture the delicate interplay between orbital configuration and spin state, as explored in Section~\ref{multiplet}.

\subsection{Multiplet calculations} \label{multiplet}
\noindent
We performed \textit{multiplet} calculations to understand the DFT results and explore their experimental consequences on STS and ESR data (Figs.~\ref{fig1}E and \ref{fig2}A). Following the methodology of Ref.~\cite{Wolf_C_and_Delgado_F_2020}, we first determined the relaxed ground-state geometry of the Ti$_\mathrm{O}^{(3)}$ adsorbate via DFT calculations (Figs.~\ref{fig5}A and B), which then served as input for the multiplet analysis. In this configuration, the three valence electrons of Ti occupy a 3$d^2$4$s^1$ shell. Here, hybridization to the 4$s$ orbital plays the role of the depolarizing hybridization found in the above DFT calculations, with the 3$d$ orbitals contain the magnetic properties of the Ti atom. By systematically varying the 4$s$–3$d$ orbital splitting $\Delta_{sd}$, the multiplet calculations reveal a crossover between the configurations 3$d^1$4$s^2$ and 3$d^2$4$s^1$ around $\Delta_{sd} \approx 1.2$~eV (Fig.~\ref{fig5}C). 

We approximate the DFT electronic structure by a 3$d^2$4$s^1$ configuration for both Ti$_\mathrm{O}^{(2)}$ and Ti$_\mathrm{O}^{(3)}$, placing these systems near the crossover regime. This suggests that the DFT-predicted $\Delta_{sd}$ for MgO$^{(2)}$ may be slightly overestimated, likely due to limitations in capturing the Ti intra-atomic correlations in the presence of external hybridization and lower symmetry. Remarkably, a single adjustable parameter—$\Delta_{sd}$—is sufficient to reproduce all experimentally observed spin states on MgO$^{(2)}$ and MgO$^{(3)}$ without invoking additional charge transfer or hydrogen adsorption/desorption mechanisms.

The resulting simulated differential conductance spectra show reasonable agreement with experiment. In particular, the calculated d$I$/d$V$ spectrum for the 3$d^1$ configuration (Fig.~\ref{fig5}D) qualitatively reproduces the features observed in Fig.~\ref{fig3}B, including the crystal-field-induced excitation from the ground-state doublet, which is known to be sensitive to tip-adsorbate distance. Likewise, for the 3$d^2$ configuration on MgO$^{(3)}$, the simulated spectrum in Fig.~\ref{fig5}E aligns with the experimental excitation near 20~mV (cf. Fig.~\ref{fig3}B).

Finally, we note that the work functions of MgO$^{(2)}$ and MgO$^{(3)}$ are nearly identical~\cite{koenig2009}, which should enable electron transfer from Ti to the Ag substrate for both cases, stabilizing a Ti$^+$ state in good agreement with the above DFT results. While previous studies have suggested hydrogenation as a cause of spin reduction,~\cite{Yang_Bae_prl_2017} our calculations show that hydrogen adsorption mainly alters the 4$s$-like density of states and does not significantly affect the 3$d$ orbital occupations responsible for the observed spin states.

\section{Conclusion}

We have combined ESR-STM spectroscopy, atom manipulation, and first-principles theory to investigate the spin and charge states of single titanium atoms adsorbed on ultrathin MgO films on Ag(100). Our experiments reveal that the spin state of Ti is strongly site-dependent: Ti adsorbed on O-atop sites of MgO$^{(3)}$  exhibits a spin $S = 1$, whereas bridge sites of MgO$^{(2)}$  and MgO$^{(3)}$, as well as O-atop sites of MgO$^{(2)}$, yield $S = 1/2$. Atom-by-atom manipulation demonstrates reversible switching between these spin states, excluding irreversible chemical modifications such as hydrogenation as the underlying mechanism.

Density functional theory and multiplet calculations support a Ti$^{+}$ charge state with three valence electrons occupying 3\emph{d} and 4\emph{s} orbitals. The observed spin-state variability is attributed to a competition between spin-polarizing and depolarizing orbital configurations, tunable by the local adsorption geometry and MgO thickness. Thermodynamic analysis rules out TiH and TiH$_3$ as stable surface species, indicating that either bare Ti or TiH$_2$ may exist depending on the availability of hydrogen, though experimental evidence points to bare Ti being dominant under typical conditions.

Altogether, our findings establish that Ti on MgO can stably host multiple spin states in a chemically pure configuration, tunable via site-selective adsorption. This opens a pathway toward the deterministic design of atomic-scale spin qubits with engineered anisotropy, coherence, and charge states, paving the way for scalable quantum nanostructures on surfaces.

\newpage
\clearpage

\section*{Methods}

\subsection{Experimental}
\noindent
An atomically clean Ag(100) substrate was prepared through repeated cycles of Ar$^+$ sputtering and thermal annealing. MgO films were grown on the Ag substrate at 580 K by evaporating Mg in an O$_2$ atmosphere of $1.1 \times 10^{-6}$ Torr. Fe and Ti atoms were deposited on the MgO surface at $< 100$ K, by precooling the sample holder in the STM state of liquid helium temperature. The STM tip was fabricated by mechanically cutting a Pt/Ir wire (0.25 mm diameter), and its apex was spin-polarized by picking up Fe atoms from the MgO surface. The tip’s spin polarization was calibrated using the asymmetry of the differential conductance (d$I$/d$V$) across the zero bias in STS spectra measured from the Ti adsorbates on the oxygen-atop sites of MgO.

STS and ESR measurements were performed on Ti adsorbates on the surfaces of both two- and three-monolayer-thick MgO layers in an ultrahigh-vacuum (UHV) chamber (base pressure $<10^{–10}$ mbar) using a $^3$He-cooled STM (Unisoku, USM1300) at $T$ = 0.4 K, equipped with two-axis superconducting magnets and high-frequency transmission cables. The ESR measurements were conducted using a commercial RF signal generator (Agilent E8257D), with the RF signal combined with a DC bias voltage through a bias tee (SigaTek, SB15D2). To read out the spin-polarized current in the ESR of the spins, a commercial current preamplifier (Femto, SPECS) followed by a lock-in amplifier (SR860, Stanford Research) was used. The RF signals were modulated at 95 Hz, with the modulation signal sent to the lock-in amplifier as a reference. The bias voltage $V_\mathrm{b}$ refers to the sample voltage while the tip is grounded. The STM constant-current feedback loop was set to open and low gain during the STS and ESR measurements, respectively.

\subsection{Theoretical}
\noindent
Multiplet calculations have been performed following the method described in Ref.~\cite{Wolf_C_and_Delgado_F_2020}.
DFT calculations have been performed using the plane-wave codes \texttt{Quantum Espresso}~\cite{QE1,QE2,QE3} and \texttt{VASP}~\cite{Kresse1996b,Kresse1999}. In addition to the PBE exchange and correlation functional, the calculations used an empirical Hubbard $U$ correction~\cite{dudarev_1998} in order to improve the description of localized $3d$ electrons. The chosen value, $U-J=2.0$~eV, is in agreement with thorough \texttt{VASP} DFT calculations~\cite{hu_choice_2011}. Van der Waals (VdW) interactions were described by the semi-empirical dispersion-forces correction D3~\cite{grimme_consistent_2010}. The calculations were performed using the projector augmented-wave method (PAW)~\cite{Blochl_1994} to treat the electron-atom potential. We used complementary calculations of the same system using \texttt{VASP} since it has been shown in literature~\cite{Jinkyung_Hyperfine_2022} that the ground state for Ti on MgO/Ag was hard to uniquely identify. Our additional calculations confirm the fundamental ground state properties of Ti on MgO/Ag, however they also show subtle differences with respect to energy splitting of the 4$s$ and 3$d$ states as well as different polarizations close to the Fermi level. This indicates that there is a dependence on the ground state solution on the pseudopotentials since all other factors (e.g. the dispersion correction) were kept consistent across different codes. For the adsorbed species, we need at least a $3\times3$ unit cell of oxygen atoms to converge the calculations with respect to cell size in order to reproduce single adsorbates. We have used four layers of Ag(100) that are critical to obtain results closer to the experiment.

\section*{Acknowledgements}
\noindent
SP, HTB, WS, YL, CW and AJH acknowledge financial support from the Institute for Basic Science (IBS-R027-D1). RR and NL thank projects  PID2021-127917NB-I00 by MCIN/AEI/10.13039/501100011033, QUAN-000021-01 by the Gipuzkoa Provincial Council, IT-1527-22 by the Basque Government, 202260I187 by CSIC, ESiM project 101046364 by the EU, and computational resources by Finisterrae III (CESGA). Views and opinions expressed are however those of the author(s) only and do not necessarily reflect those of the EU. Neither the EU nor the granting authority can be held responsible for them.  CW thanks Fernando Delgado and Susanne Baumann for insightful discussions.

\section*{Author Contribution}
\noindent
SP conceived the experiment. SP, HTB, WS, and YL performed experiments.
NL and RR performed the VASP DFT calculations. CW and CL performed QE DFT and multiplet calculations.
All authors discussed the results and prepared the manuscript.

\bibliography{references}

\clearpage
\newpage

\begin{figure} [t!]
\centering
\includegraphics[width = 9cm]{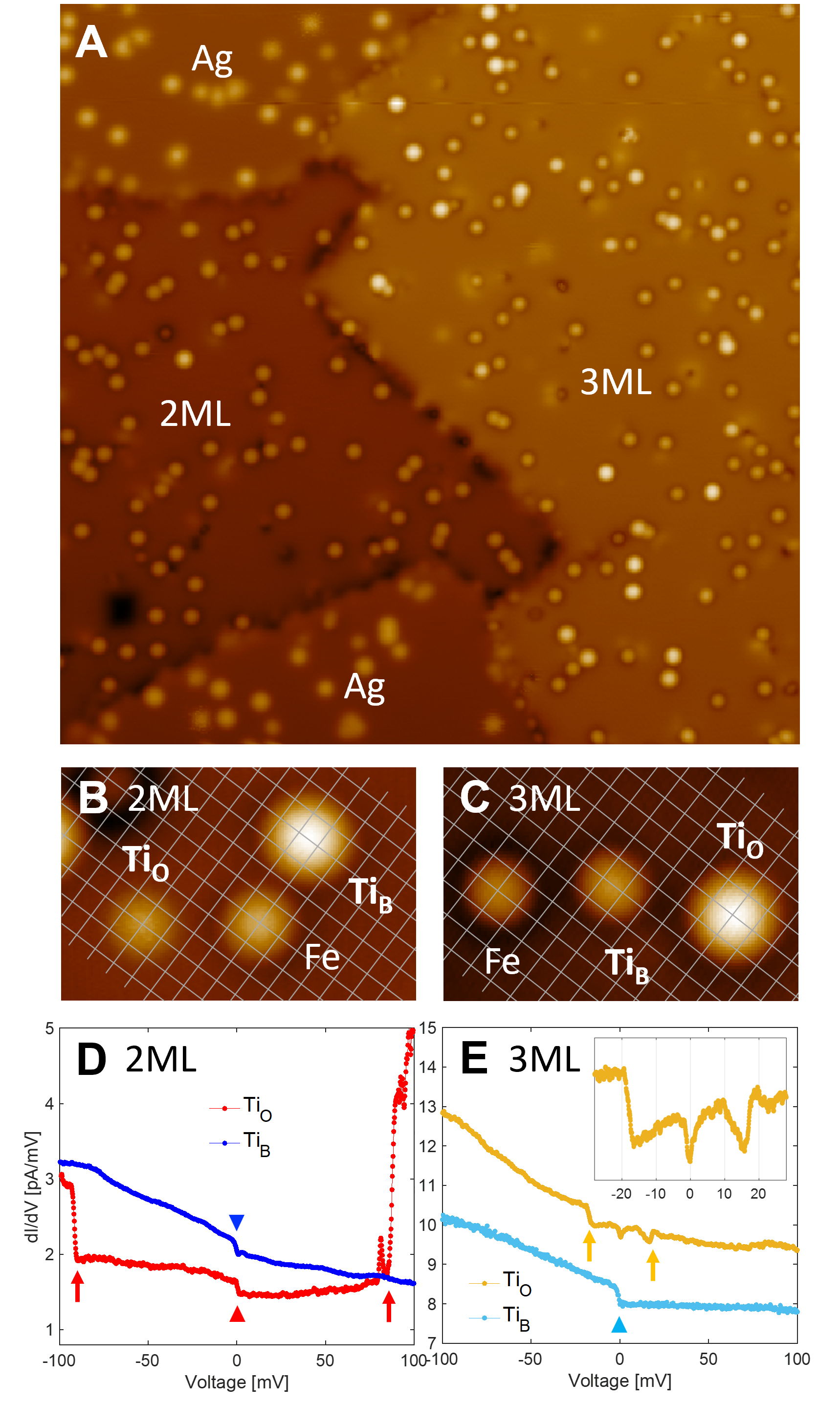}
\caption{\textbf{Atomic species on MgO/Ag:} (A) An $50\times50$ nm$^2$ STM topographic image of the sample surface (B and C) High resolution STM images of 2 ML and 3 ML MgO surfaces with Ti and Fe adsorbates on oxygen-atop sites. Set points: $V_{\rm b} = 100$ mV and $I_{\rm tun} = 10$ pA. The overlayed gray meshes indicate the MgO lattice of which each crossing point is the oxygen-atop site. (D and E) STS spectra measured on Ti atoms. Arrows indicate characteristic features stemming from tunneling via inelastic channels. Triangles point the features induced by the spin-polarized tips. The inset in (E) is a zoom-in of the spectrum around zero bias. Set points: $V_{\rm b} = 100$ mV and $I_{\rm tun} = 100$ pA.}
\label{fig1}
\end{figure}
\clearpage
\newpage

\begin{figure} [t!]
\centering
\includegraphics[width = 10cm]{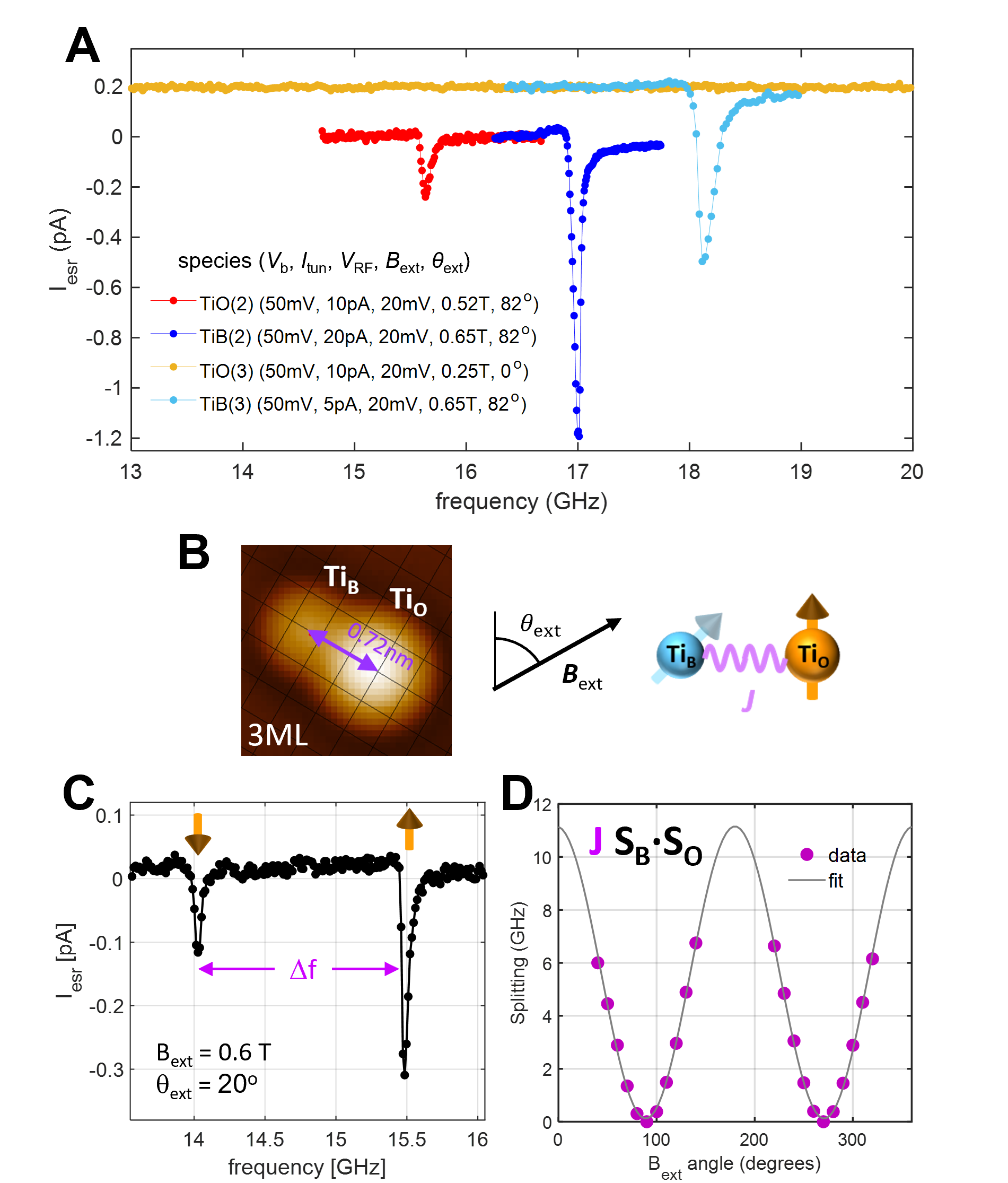}
\caption{\textbf{ESR of Ti adsorbates on MgO:} (A) ESR spectra measured on the four Ti species. The spectra measured on 3ML are shifted by 0.2 pA for clarity. Measurement conditions for each spectrum are denoted in the legend. (B) STM image of a Ti$_{\rm B}^{(3)}$ with a nearby Ti$_{\rm O}^{(3)}$ at a distance of 0.72 nm, with a depiction of spin-spin coupling ($J$) and external magnetic field ($B_{\rm ext}$). The overlayed square mesh indicates the MgO lattice of which each crossing point is the oxygen-atop site. (C) An ESR spectrum measured on the Ti$_{\rm B}^{(3)}$ coupled with a Ti$_{\rm O}^{(3)}$, as shown in B. (D) Field angle dependence of the splitting ($\Delta f$) measured from ESR spectra. Gray curve is a fit of $\Delta f$ to $\cos(\theta_{\rm ext})$. The angle of the external field ($\theta_{\rm ext}$) is depicted in B.
%(E) Multiplet calculations performed to 3$d$ electron orbital states of Ti adsorbed on an O-top of 3ML MgO surface ($B_{\rm ext}$ = 0.6 T, $\theta_{\rm ext} = 20^\circ$). The expectation values of the spin and orbital angular momenta, $S_{\rm z}$ and $L_{\rm z}$, are given on the right side of each eigenstate.
}
\label{fig2}
\end{figure}
\clearpage
\newpage

\begin{figure} [t!]
\centering
\includegraphics[width = 9cm]{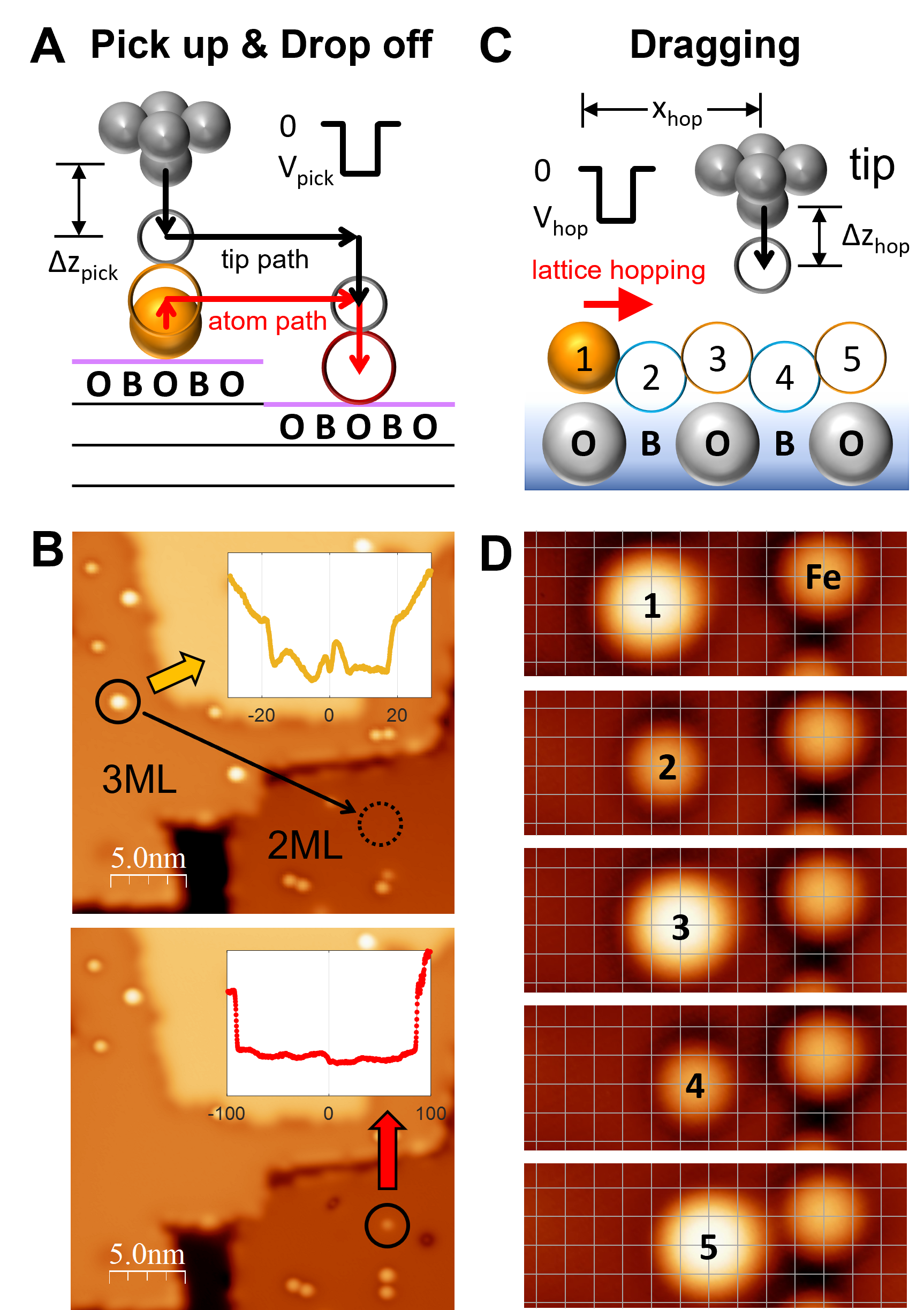}  
\caption{\textbf{Manipulations of Ti atoms on 3ML MgO:} (A) Schematic describing moving a Ti atom by pick-up and subsequent drop-off processes using a STM tip. (B) STM images before (upper) and after (lower) moving a Ti atom from a 3ML to a 2ML MgO patches, respectively, as depicted by the black arrow. The insets are the STS spectra measured before and after moving the atom. (C) Schematic describing hopping of a Ti atom on 3ML MgO using a DC bias voltage pulse. (D) 5 STM images sequentially measured before (1) and after each hopping of a Ti atom from positions 1 to 5, as described in C. The STM images in B and D are measured at $V_{\rm b} = 100 $ mV and $I_{\rm tun} = 10 $ pA.}
\label{fig3}
\end{figure}
\clearpage
\newpage

\begin{figure}
\centering
\includegraphics[width = 1.\textwidth]{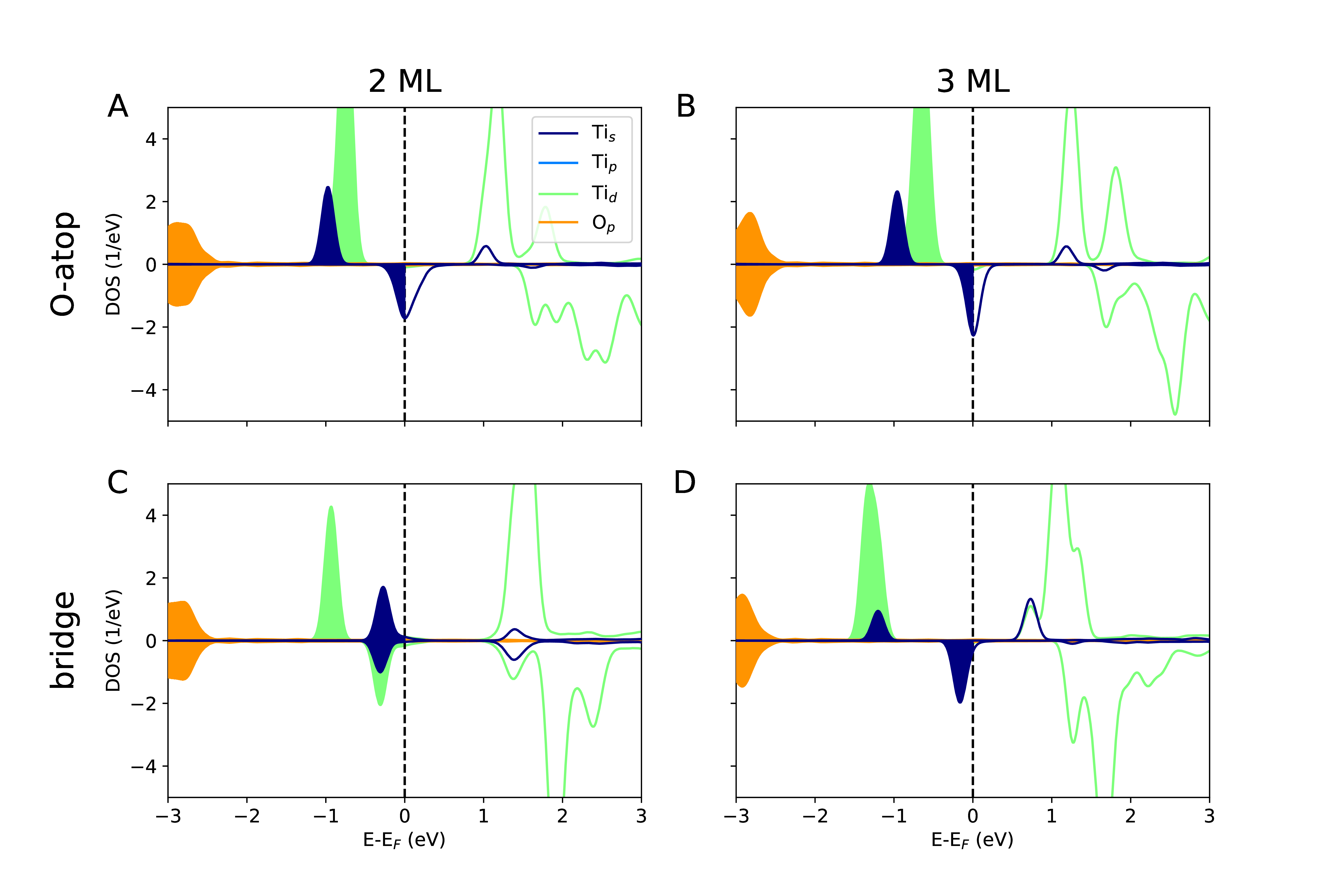}  
\caption{\textbf{PDOS of Ti on MgO/Ag (100).} The density of states projected on $s,p,d$ orbitals of the Ti atom and the nearest oxygen for four cases that have been displayed in columns depending on the number of monolayers (ML) of MgO and in rows for the oxygen-atop ($O$) and bridge ($B$) sites: (A)Ti$_O$(2), (B) Ti$_O$(3), (C) Ti$_B$(2), (D) Ti$_B$(3).}\label{PDOS_Ti}
\end{figure}
\clearpage
\newpage

\begin{figure} [t!]
\centering
\includegraphics[width = 0.8\textwidth]{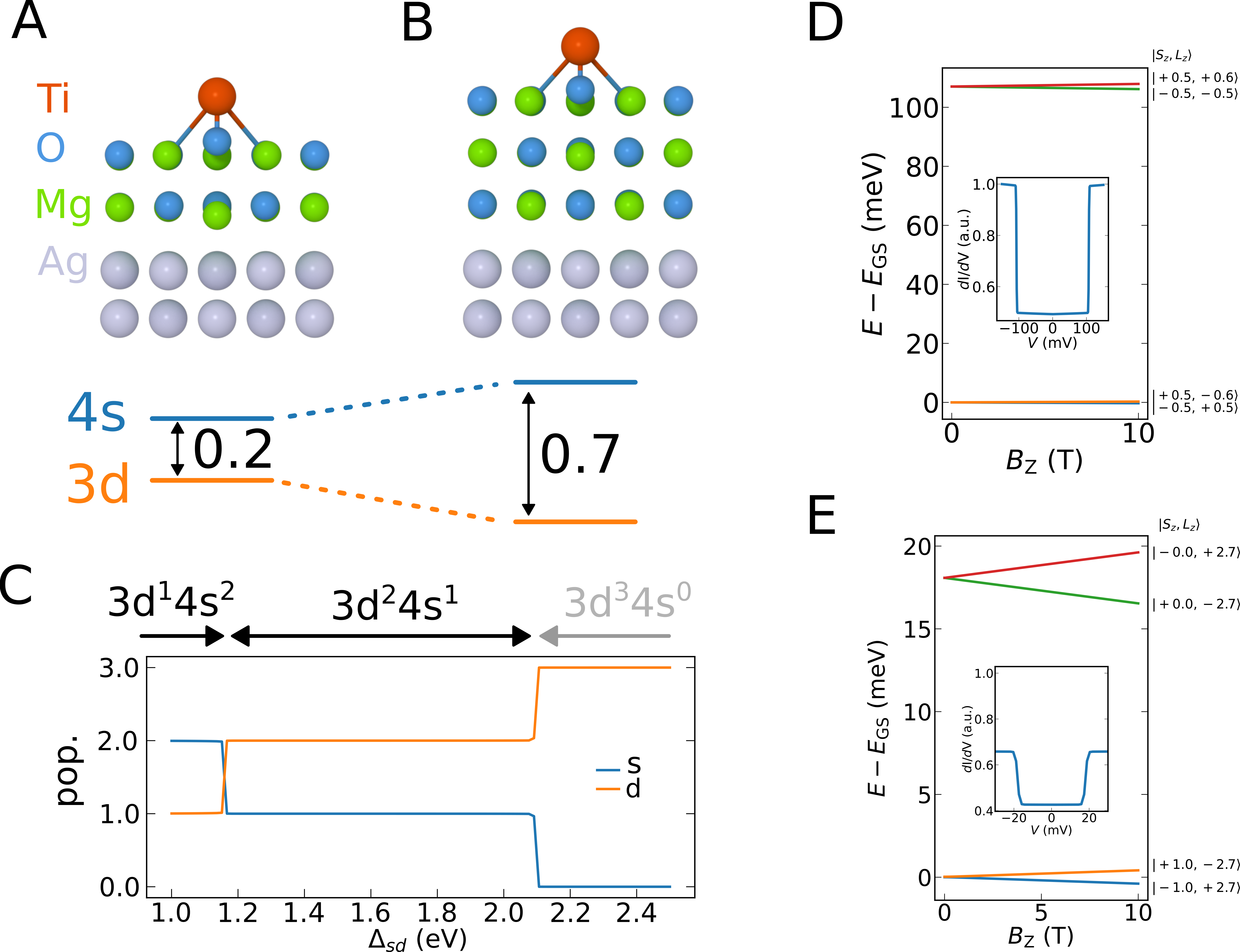}  
\caption{\textbf{Multiplet calculations Ti on MgO$^{(2)}$ and MgO$^{(3)}$ on Ag (100).} (A) and (B) show the relaxed geometry of Ti on  MgO$^{(2)}$ and MgO$^{(3)}$. When going from 2 to 3 ML of MgO the $s-d$ splitting $\Delta_{sd}$ increases, as shown schematically below. (C) spin states of the Ti$^+$ as function of $\Delta_{sd}$. Indicated are three possible orbital configurations for the same charge state. (D) multiplet calculations and simulated $dI/dV$ spectrum for the case of 3$d^1$, showing two strong excitations around 100 meV. (E) shows the same calculation for a 3$d^2$ orbital configuration, which shows excitations close to 20 meV, in reasonable agreement with the experiment (see inset in Fig.~\ref{fig3}B). }
\label{fig5}
\end{figure}

%Supplementary
\clearpage
\pagebreak
\setcounter{page}{1}
\setcounter{equation}{0}
\setcounter{figure}{0}
\renewcommand{\theequation}{S.\arabic{equation}}
\renewcommand{\thefigure}{S\arabic{figure}}
\renewcommand*{\thepage}{S\arabic{page}}
\onecolumngrid
\begin{center}
{\large \textbf{Supplementary Information for \\ ``\ourtitle"}}\\
\end{center}
\tableofcontents
\section{Further experimental analysis}

\subsection{STM topographic images of the sample surface}
In Fig.~\ref{fig_topo_bias}, we present the STM topographic images of the as-prepared MgO/Ag surface measured at various bias voltages ($V_{\rm b}$), where we observed that the apparent height of MgO patches changes with $V_{\rm b}$. The conduction band edge of the ultrathin MgO on Ag(100) is known to set in between 2 and 3 eV above the Fermi energy~\cite{schintke2001insulator}. In order to better visualize the geometrical height of the MgO patches in the STM images, we subtract the topographic contrast for $V_{\rm b}$ = 2 V from that for $V_{\rm b}$ = 3 V, as shown in Fig.~\ref{fig_topo_bias}(E). This is also a way to effectively get rid of the contribution of the Ag terrace heights to the STM apparent topographic contrast. It is clearly observed that all the MgO patches in the field of view have the same geometric heights, which are revealed to be the 2ML-thick by the current vs. tip height measurements. Of more interesting is that the height of the patches are spatially uniform down to the atomic scale except for a very low density of point defects.

%\subsection{Work function measurements on the sample surface}
%The intercalation of oxygen at the MgO-Ag interface can drastically changes the electronic properties of MgO such as the work function, as predicted by the DFT study (see Fig.~\ref{fig_WF}(B)). Work function of a surface can be extracted from the current vs. tip height measurement $I(z_{\rm tip})$ in a STM~\cite{qi2007atomic}. Figure~\ref{fig_WF}(A) shows $I(z_{\rm tip})$ measurements performed for the three surfaces of bare Ag(100), 2ML, and 3ML MgO/Ag(100), used in this work. The extracted work functions are 4.57, 3.32, and 3.04 eV for Ag(100), 2ML, and 3ML MgO/Ag(100), respectively. The work function of Ag(100) is consistent with previous results (4.64 eV)~\cite{dweydari1975work} and different from the DFT calculation (Fig.~\ref{fig_WF}(B)) by about 0.4 eV. Taking into consideration of this difference, the measured work function of 2ML MgO agrees with the case of no oxygen intercalation as shown in Fig.Fig.~\ref{fig_WF}(B). In addition, inhomogeneity in the STM topography of MgO are not observed except for some point defects in Fig. 1 and Fig.~\ref{fig_topo_bias}, which indicates either no oxygen or an oxygen intercalation of very atomic uniformity at the interface. In summary, the results from work function measurement together with the uniform STM topographic contrast of the MgO patches suggest no intercalation of oxygen at the interface.

\subsection{Work function measurements on the sample surface}
\begin{figure*} [t!]
\centering
\includegraphics[width = 0.8\textwidth]{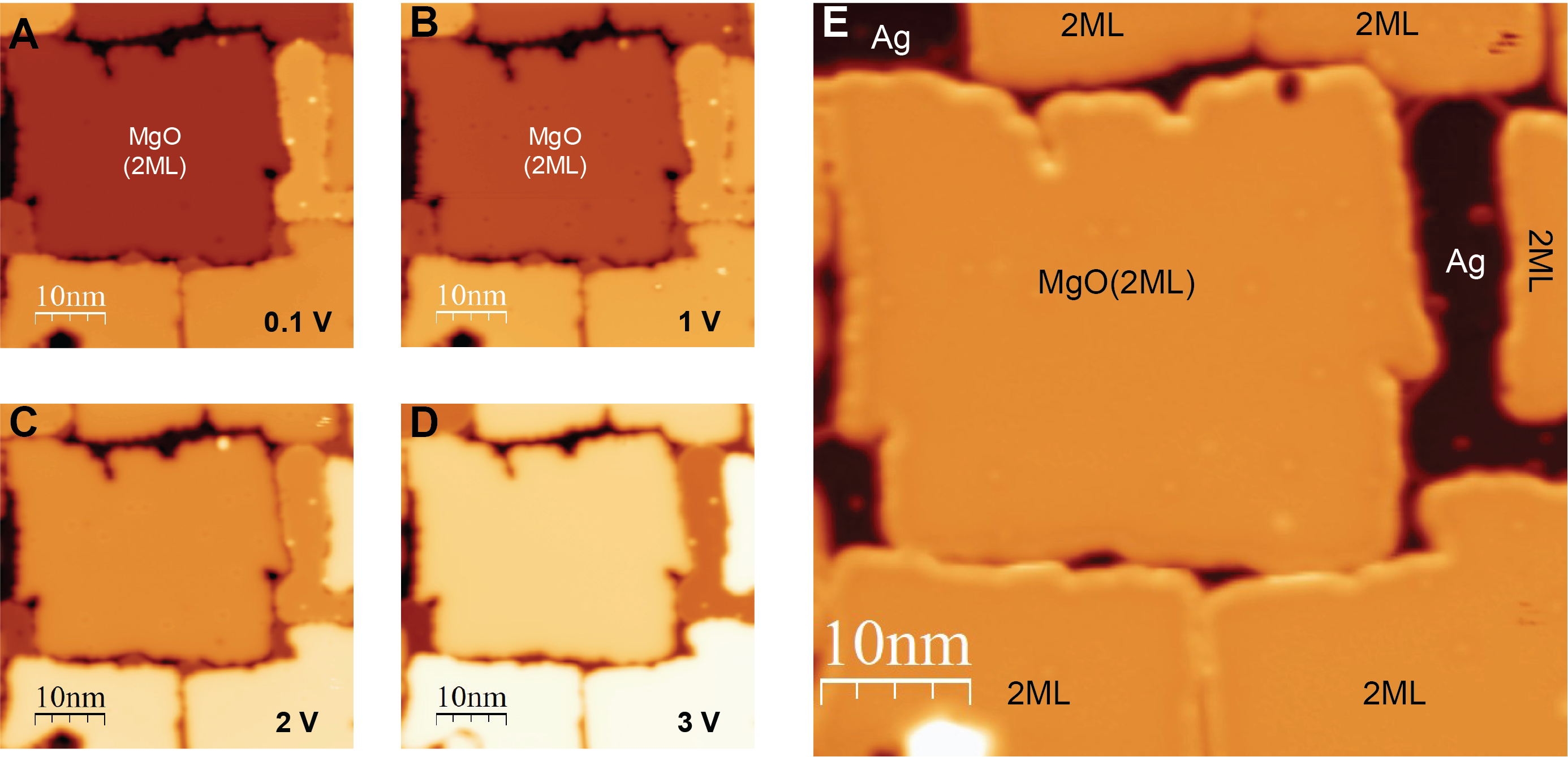}  
\caption{\textbf{STM topographic images of a region covered mainly by 2ML MgO at various bias voltages ($V_b$):} (A) $V_{\rm b}$ = 0.1 V. (B) $V_{\rm b}$ = 1 V. (C) $V_{\rm b}$ = 2 V. (D) $V_{\rm b}$ = 3 V. (E) Topographic contrast of the image in D subtracted by that in C.}
\label{fig_topo_bias}
\end{figure*}

Oxygen intercalation at the MgO–Ag interface can significantly alter the electronic properties of MgO, including its work function, as predicted by DFT calculations (see Fig.\ref{fig_WF}(B)). We use STM current versus tip-height measurements, $I(z_{\rm tip})$, to estimate the work function of pristine surface\cite{qi2007atomic}. Figure~\ref{fig_WF}(A) displays $I(z_{\rm tip})$ curves acquired for three surfaces studied in this work: bare Ag(100), and 2ML and 3ML MgO/Ag(100). The extracted work function values are 4.57 eV for Ag(100), 3.32 eV for 2ML MgO, and 3.04 eV for 3ML MgO. The measured work function for Ag(100) agrees well with previous experimental reports (4.64 eV)\cite{dweydari1975work}, although it differs by approximately 0.4 eV from the DFT results (see Fig.\ref{fig_WF}(B)). Accounting for this discrepancy, the experimentally measured work function for 2ML MgO is consistent with the DFT prediction in the absence of oxygen intercalation~\cite{MgOWorkFunction}, as indicated in Fig.~\ref{fig_WF}(B). The trend with MgO thickness is also nicely recover in DFT calculations~\cite{Wolf_C_and_Delgado_F_2020}. Furthermore, the STM topography of MgO shows no significant inhomogeneity apart from a few point defects (see Fig.1 and Fig.\ref{fig_topo_bias}), suggesting either the absence of oxygen intercalation or an extremely uniform distribution of intercalated oxygen at the atomic scale. In summary, the combination of work function measurements and uniform STM topographic contrast strongly indicates that there is no oxygen intercalation at the MgO/Ag interface.

\begin{figure*} [h!]
\centering
\includegraphics[width = 0.9\textwidth]{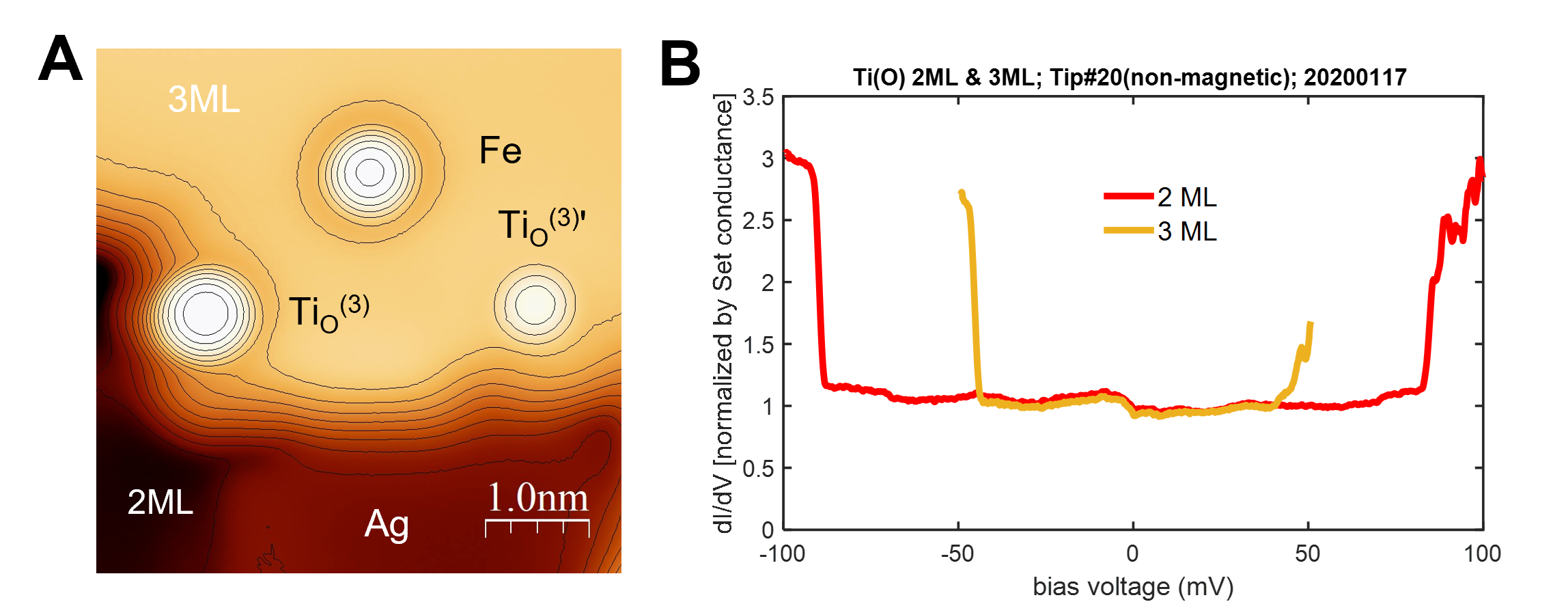}  
\caption{\textbf{Another Ti species adsorbed on an oxygen site of 3ML MgO, Ti$_{\rm O}^{(3)}$'} (A) STM topographic image of Ti and Fe adsorbates on an 3ML MgO. $V_{\rm b}$ = 50 mV. $I_{\rm tun}$ = 10 pA. (B) STS spectrum measured on the Ti$_{\rm O}^{(3)}$' in A (orange). A spectrum measured on a Ti$_{\rm O}^{(2)}$ (red) using the same tip is shown together for comparison.}
\label{fig_TiO3ml_small}
\end{figure*}

\clearpage

\begin{figure*} 
\centering
\includegraphics[width = 0.9\textwidth]{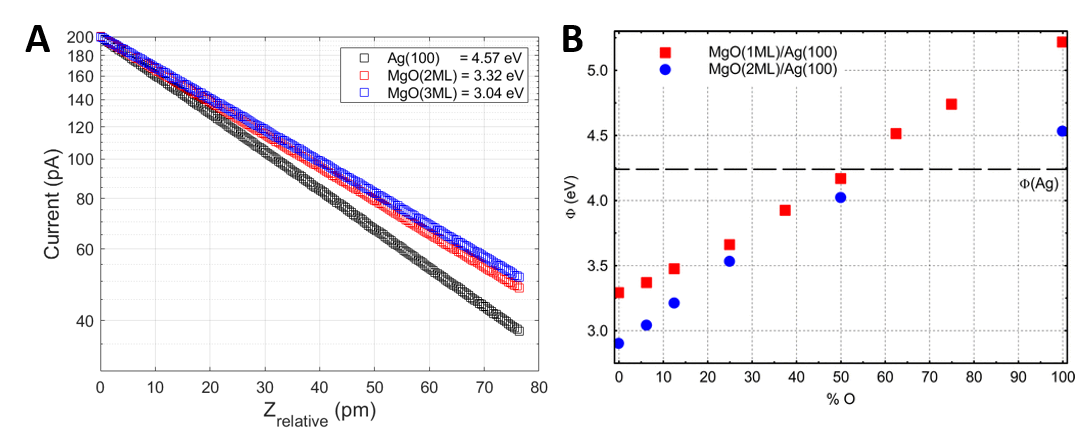}  
\caption{\textbf{Current vs. tip height ($z_{\rm relative}$):} (A) Measured work function of Ag(100), 2ML and 3ML MgO/Ag(100) in this work. (B) DFT results from Ref.~\onlinecite{MgOWorkFunction} on the dependence of work function for 1ML and 2ML MgO/Ag(100) as a function of intercalated oxygen concentration. Figure (B) was reprinted from the reference~\cite{MgOWorkFunction}.}
\label{fig_WF}
\end{figure*}

%\newpage
\clearpage

\section{Experimental parameters for atom manipulations}

As shown in the section II.C, we use two distinct types of atom manipulation techniques to build artificial multiple spin structures on a MgO surface. Below presented are typical experimental protocols for the two methods. In Tables~\ref{Table_1} and ~\ref{Table_2}, the experimental parameters used for \textit{Pick up and Drop off} and \textit{Dragging} of a Ti atom on a 3ML MgO surface. We use both methods for a Ti$_{\rm O}^{(3)}$ atom, however, only \textit{Dragging} works for a Ti$_{\rm B}^{(3)}$ atom.\\

%\noindent
\textbf{Pick up \& Drop off:}\\
1. move the tip right on the target atom\\
2. set initial tip height by choosing tunneling conditions $I_{0}$ and $V_{0}$\\
3. switch off the tunneling Feedback circuit\\
4. move the tip down by $\Delta z_{\rm pick}$\\
5. apply bias voltage pulse $V_{\rm pick}$ for $\tau_{\rm pick}$\\
6. move the tip up by $\sim 1$ nm and move laterally to the new position of the atom\\
7. move down the tip by $\Delta z_{\rm drop}$\\
8. apply bias voltage pulse $V_{\rm drop}$ for $\tau_{\rm drop}$\\
9. lift up the tip by $-\Delta z_{\rm pick}$\\
10. set the tunneling conditions back to $I_{0}$ and $V_{0}$\\
11. switch on the tunneling Feedback circuit\\

%\noindent
\textbf{Dragging:}\\
1. move the tip at the position $x_{\rm hop}$ away from the target atom\\
2. set initial tip height by choosing tunneling conditions $I_{0}$ and $V_{0}$\\
3. switch off the tunneling Feedback circuit\\
4. move the tip down by $\Delta z_{\rm hop}$\\
5. apply bias voltage pulse $V_{\rm hop}$ for $\tau_{\rm hop}$\\
6. lift up the tip by $-\Delta z_{\rm hop}$\\
7. set the tunneling conditions back to $I_{0}$ and $V_{0}$\\
8. switch on the tunneling Feedback circuit\\

%\newpage
%\clearpage

\begin{table*}%[b!]
\centering
\footnotesize
\begin{tabular}{c| c| c| c| c| c| c| c}
\hline% \hline
 $I_{\rm 0}$ (pA) & $V_{\rm 0}$ (mV) & $\Delta z_{\rm pick}$ (pm) & $V_{\rm pick}$ (mV) & $\tau_{\rm pick}$ (ms) & $\Delta z_{\rm drop}$ (pm) & $V_{\rm drop}$ (mV) & $\tau_{\rm drop}$ (ms) \\  \hline 
10 & 100 & 200 & 550 & 200 & 500 & $-$10 & 200 \\  \hline 

% TiH$_{\rm O}^{(2)}$ & 2.07 & 2.24  & $-$0.95 \\  \hline
% TiH$_{\rm B}^{(3)}$ & 2.09 & 1.86 & $-$1.17  \\ \hline 
% TiH$_{\rm O}^{(3)}$ & 1.82 & 2.41 & $-$0.85\\ }\hline
% \hline
\end{tabular}
\caption{Experimental parameters for \textit{Pick up and Drop off} a Ti$_{\rm O}^{(3)}$ atom.}
\label{Table_1}
\end{table*}

\begin{table*}%[b!]
\centering
\begin{tabular}{c|| c| c| c| c| c| c}
\hline% \hline
Species & $I_{\rm 0}$ (pA) & $V_{\rm 0}$ (mV) & $x_{\rm hop}$ ($a_{\rm O-O}$) & $\Delta z_{\rm hop}$ (pm) & $V_{\rm hop}$ (mV) & $\tau_{\rm hop}$ (ms) \\  \hline 
Ti$_{\rm O}^{(3)}$ & 10 & 100 & 1.5 & 300 & $-$300 & 200 \\  \hline
Ti$_{\rm B}^{(3)}$ & 10 & 100 & 1.5 & 300 & 300 & 200 \\  \hline
% TiH$_{\rm O}^{(2)}$ & 2.07 & 2.24  & $-$0.95 \\  \hline
% TiH$_{\rm B}^{(3)}$ & 2.09 & 1.86 & $-$1.17  \\ \hline 
% TiH$_{\rm O}^{(3)}$ & 1.82 & 2.41 & $-$0.85\\ \hline
% \hline
\end{tabular}
\caption{Experimental parameters for \textit{Dragging} single Ti atoms on a 3ML MgO. The $a_{\rm O-O}$ is the nearest neighbor distance between two oxygen atoms on the MgO surface.}
\label{Table_2}
\end{table*}

\newpage
\clearpage

\section{DFT calculations}
The calculations have been performed using the  VASP code \cite{Kresse1996b} with the projector augmented-wave (PAW) method \cite{Kresse1999}  and the \texttt{Quantum Espresso} code~\cite{QE1,QE2,QE3}. Both codes are based on a plane-wave expansion of the wave functions. We will first discuss the results obtained with VASP, for details of the QE calculations see the next section. Both codes give a coherent picture with small numerical deviations contributing to show the reliability of our results.

\subsection{VASP-code results: Electronic structure and hydrogenation}

The basis was expanded with an energy cut-off of 400 eV. The exchange and correlation functional is the PBE-GGA one \cite{PBE}. This functional was completed with van der Waals interactions introduced through the D3 Grimme scheme~\cite{grimme_consistent_2010}. The unit cell in all calculations was taken to be a $3\times3$ network of O atoms (and Mg atoms). The lattice parameter was the one of bulk Ag self-consistently obtained using van der Waals corrections. The slabs were made by two and three ML of MgO on four layers of Ag(100), except in one case with eight layers to check for the convergence of Ti$_{\rm O}$ (2). The calculations were converged using $3\times3\times1$ k-point sampling.

In the following, we present DFT total energy calculations of TiH$_x$, with $x=0,1,2$, in both gas phase and adsorbed phases on MgO/Ag(100). Our results points towards Ti$^+$ as the most likely state of the adsorbate based on total energy considerations. In addition, we present multiplet calculations for 2 and 3 ML of MgO to explain the change in the measured STS spectra.

\subsection{TiH$_x$ in gas phase}
\noindent
DFT calculations for the gas-phase Ti recover its known electronic configuration [Ar]$3d^24s^2$, leading to a $S=1$ ground state. Ti presents a large affinity for H atom given its open shell structure,\cite{Brown1988, natterer2013} and easily forms TiH$_x$ for $x$ all the way from 0 to 4. We can picture the different molecules as starting in $sd$, $sd^2$ and $sd^3$ hybrid orbitals that saturate as H atoms are added. In this way, we can expect a spin $S=3/2$ for TiH since the H will close one shell and three will be left half-occupied. For TiH$_2$ we find a spin $S=1$ for the same reason but now with 2 H atoms.  TiH$_3$ is then a spin $S=1/2$ and TiH$_4$ (titanium tetrahydride) is a closed-shell tetrahedral molecule. Of all these molecules, TiH$_4$ is the most stable one, followed by TiH$_2$.

\subsection{TiH$_x$ on MgO/Ag(100): Total energy}
\noindent
From the above gas-phase considerations, we can expect strong bonds of TiH$_x$ to the MgO/Ag(100) surface for $x$ = $0-2$. TiH$_2$ can easily coordinate its two dangling bonds both on O-top and bridge sites, also for TiH and Ti. However, it is very difficult to converge calculations for TiH$_3$ on O-top site of MgO/Ag(100) despite the possibility of forming a Ti-O directional bond. Rather, it easily loses one H onto the surface, forming TiH$_2$ adsorbate on MgO as discussed in the following.
%indicating that two H are likely to attach to an adsorbed Ti atom but not three.

%\subsubsection*{Gibbs free energy}
%\noindent
\textbf{Gibbs Free Energy:} When comparing the adsorption of different species, the Gibbs free energy ($G=H-TS$, $H$ being the enthalpy and $TS$ the product of temperature and entropy) determines the prevalence of one species over another~\cite{Reuter}. Calculation of $G$ requires the total energies of the different species, where a key component is the chemical potential of the hydrogen source.
Assuming that the source of hydrogen is its molecular form (H$_2$), which is energetically favorable even at the experimental temperature ($\sim 0.4$ K),
the chemical potential per H atom ($\mu_{\rm H}$) is given by the energy required to dissociate a H$_2$ molecule, divided by 2, i.e. $\mu_{\rm H}=E[\text{H}]=E^\mathrm{(diss)}[\text{H}_2]/2$. Here we take the energy reference from gas-phase H$_2$, then contribution of any other source of hydrogen ($\text{H}_{\text{new}}$) can be considered by using its chemical potential relative to a single hydrogen in the gas phase:
%\[
\begin{equation}
\Delta \mu = E[\text{H}_{\text{new}}]-E[\text{H}].
\end{equation}
%\]
Next, the chemical potential $\mu_x$ of one hydrogen in a TiH$_x$ adsorbate is calculated:
\begin{equation}
    \mu_x = E[\text{TiH}_x]-(E[\text{TiH}_{x-1}]+E[\text{H}]),
\end{equation}
which corresponds to the binding energy of a hydrogen atom in the TiH$_x$ adsorbate. Here, $E[\text{TiH}_x]$ is the total energy of the TiH$_x$ on MgO/Ag(100) and $E[\text{H}]$ is the energy per hydrogen atom.
Following this discussion, the change in Gibbs free energy $\Delta G$ for the adsorption of $x$ hydrogen atoms on Ti is given by:
\begin{equation}
    \Delta G (\Delta\mu) = x \mu_{x} - x\Delta\mu.
    \label{GibbsEq}
\end{equation}
\noindent Here, we assume that other sources of free energy, such as molecular rotations, vapor pressure, vibrations, etc., are negligible compared to the strength of the chemical bonds at the experimental temperature ($\sim 0.4$ K). Thus, plots of $\Delta G (\Delta\mu)$ allows us to explore the stability of different TiH$_x$ species with respect to various hydrogen sources. 

\begin{figure} [t!]
\centering
\includegraphics[width = 10cm]{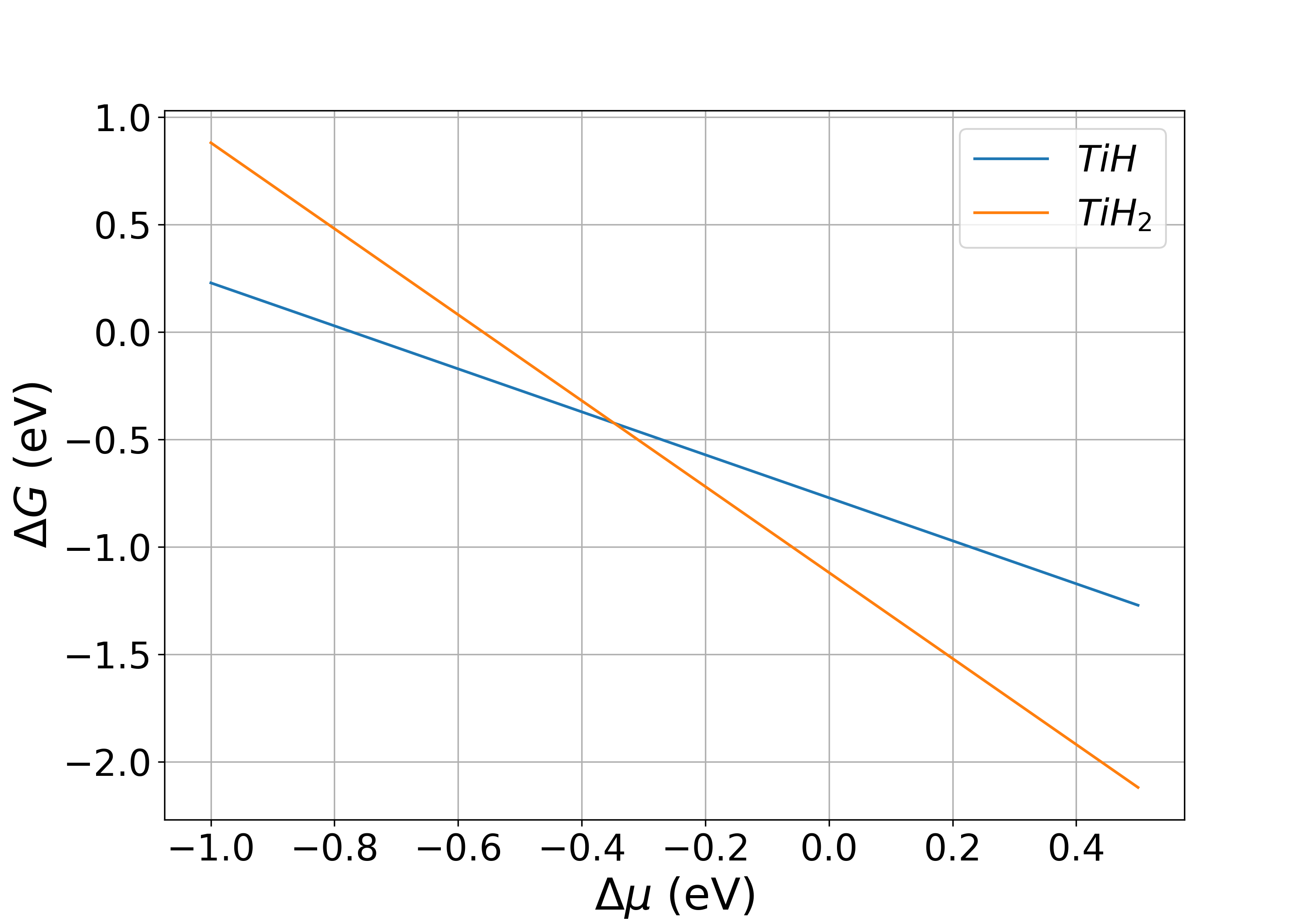}  
\caption{\textbf{Gibbs free energy:} DFT calculations of the Gibbs free energy of TiH$_x$  ($x=0,\, 1,\,2$) adsorbed on the oxygen-bridge site of 2 ML of MgO on Ag (100) as a function of Hydrogen chemical potential. The graphs shows several regions where different TiH$_x$ species can stabilize depending on the chemical potential of the source of H atoms. The chemical potential is referred to the gas-phase H$_2$ molecule such that zero corresponds to a H source of gas-phase H$_2$. The Gibbs free energy is referred to $x=0$. Thus, if the source of H is the gas-phase H$_2$ in the chamber, we find that the $\Delta G (\Delta \mu=0)$ is more negative for $x=2$, and the calculation predicts that the more stable phase corresponds to adsorbed TiH$_2$. The values of $\Delta \mu$ that correspond to $\Delta G=0$ give the H binding energy  to TiH and to TiH$_2$ per H, respectively.}
\label{figS4}
\end{figure}

Figure~\ref{figS4} shows calculated $G$ as a function of $\Delta\mu$ for TiH and TiH$_2$, with adsorbed atomic Ti and gas-phase H$_2$ as references on the bridge site. It is evident that $G$ is lower for a TiH$_2$ adsorbate than TiH when the source of hydrogen is gas-phase H$_2$ ($\Delta\mu=0$).

Without loss of generality, we assume that hydrogen first adsorbs on the MgO surface and then diffuses until it encounters the adsorbed Ti atoms. This matches the anecdotal experimental observation that samples of Ti on MgO/Ag slowly degrade over time, if TiH is formed (TiH is never $S=1/2$, see Table~\ref{Table_3}). 

Only clusters or nanostructures of Ti may act as a stronger hydrogen getter~\cite{H-Ti_chain}, which are unlikely in our case because the experiments were performed for a very low coverage ($\lesssim 1~\%$) on the surface, with well isolated single Ti adsorbates (see Fig.~\ref{fig1}A). The H chemical potential will likely be close to zero, corresponding to the reference of gas-phase H$_2$, since it is difficult to bind H much stronger than in its own molecule. In this case, Fig.~\ref{figS4} shows that 
TiH$_2$ prevails on MgO/Ag (100).

In conclusion, the presence of TiH$_2$ will depend on the relative abundance of hydrogen and its source.  In an abundance of hydrogen, TiH$_2$ will be the thermodynamically stable species. However, if hydrogen is scarce, together with the difficulty to adsorb and diffuse atomic and molecular hydrogen on MgO~\cite{Wu}, Ti will be found on the MgO surface. 

\clearpage
\newpage

\textbf{TiH}: Previous DFT calculations that did not include the silver substrate~\cite{Yang_Bae_prl_2017,Willke_Bae_science_2018} have typically used TiH as adsorbed species. 
Our DFT calculations consistently predict an $S=1$ state for all evaluated conformations of TiH ($x = 1$) on MgO/Ag(100), as shown in Table~\ref{Table_3}. However, from both thermodynamic and magnetic perspectives, DFT clearly indicates that TiH is unlikely to be the majority adsorbate on MgO/Ag(100). Ti cedes a full electron, keeping a similar configuration for the four cases studied here. These results are clearly at odds with all the experimental evidence.
\begin{table*}%[t!]
\centering
\begin{tabular}{| c| c| c| }
\hline \hline
& Chem. energy (eV) & Ti magnetic moment ($\mu_{\rm B}$)  \\ 
 \hline
 TiH$_{\rm B}^{(2)}$ & 2.53 & 1.86 \\  \hline

 TiH$_{\rm O}^{(2)}$ & 2.07 & 2.24   \\  \hline
 TiH$_{\rm B}^{(3)}$ & 2.09 & 1.86    \\ \hline 
 TiH$_{\rm O}^{(3)}$ & 1.82 & 2.41  \\
 \hline
 \hline
\end{tabular}
\caption{\textbf{DFT results for TiH on two and three layers of MgO on Ag(100):} The first column contains the system. The second column is the chemical energy in eV defined as the energy needed to desorb the TiH system into atomic Ti and H. The magnetic moment from Bader charges~\cite{Tang_2009} is given in the third column.  The present results show a magnetic moment approaching 2$\mu_{\rm B}$ in all cases, pointing at a $S=1$ configuration for TiH on 2 and 3 ML of MgO on Ag (100).}
\label{Table_3}
\end{table*}

\textbf{TiH$_2$}: We find that TiH$_{2}$ adsorbed at bridge sites of both 2 and 3 ML MgO exhibit characteristics of nearly perfect spin 1/2 systems, with magnetic moments of 0.97 $\mu_{\rm B}$ (see Table~\ref{Table_4}). 
\begin{table*}
\centering
\begin{tabular}{| c| c| c| }
\hline \hline
& Chem. energy (eV) & Ti magnetic moment ($\mu_{\rm B}$)  \\ 
 \hline
 TiH$_{\rm 2,B}^{(2)}$ & 2.97 & 0.97 \\  \hline

 TiH$_{\rm 2,O}^{(2)}$ & 2.78 & 1.90   \\  \hline

 TiH$_{\rm 2,B}^{(3)}$ & 2.51 & 0.97 \\  \hline

 TiH$_{\rm 2,O}^{(3)}$ & 2.56 & 1.90 \\
 \hline
 \hline
\end{tabular}
\caption{\textbf{DFT results for TiH$_2$ on two and three layers of MgO on Ag(100):} The first column contains the system. The second column is the chemical energy in eV as before, where you remove TiH$_2$ from the surface and separate it in Ti and atomic H. The magnetic moment for the Ti  atom calculated from Bader charges is given in Bohr magnetons ($\mu_{\rm B}$), third column. }
\label{Table_4}
\end{table*}

\textbf{Ti}: As shown in Table~\ref{Table_5}, Ti$_{\rm B}$$^{(2)}$ and Ti$_{\rm B}$$^{(3)}$ exhibit $S=1/2$ configurations, while Ti$_{\rm O}$$^{(2)}$ and Ti$_{\rm O}$$^{(3)}$ show $S=1$ configurations. As in the cases of TiH$_2$, these results qualitatively agree with experimental observations. In both cases, Ti and TiH$_2$, the predicted magnetic moment fits a $S=1$ configuration contrary to the experiment. We believe this is due to the closeness of a depolarizing transition that is difficult to capture at this level of theory, as discussed in the main text. 

\begin{table*}
\centering
\begin{tabular}{| c| c| c| }
\hline \hline
& Chem. energy (eV) & Ti magnetic moment ($\mu_{\rm B}$) \\ 
 \hline
 Ti$_{\rm B}^{(2)}$ & 1.76 & 1.31  \\  \hline

 Ti$_{\rm O}^{(2)}$ & 1.58 & 2.28   \\  \hline

 Ti$_{\rm B}^{(3)}$ & 1.23 & 1.09  \\  \hline

 Ti$_{\rm O}^{(3)}$ & 1.40 & 2.10  \\
 \hline
 \hline
\end{tabular}
\caption{\textbf{DFT results for Ti on MgO/Ag(100):} The first column contains the system. The chemical energy is defined as the energy needed to desorb the Ti atom. The magnetic moment (third column) points at Ti on the bridge site as a $S=1/2$ species, while on top of an O atom, Ti behaves as a $S=1$ atom.}
\label{Table_5}
\end{table*}

Experimentally, the bridge adsorption site is favorable for the two-monolayer MgO/Ag(100) system, while the top site is the favored one on three monolayers. The computed chemisorbed energies only support this trend for atomic Ti adsorbates with significative energy differences, Table~\ref{Table_5}. This is in agreement with the assignment of Ti as the species object of the above experiments.

\section{QE DFT calculations: Multiplet Calculations}
\subsection{DFT calculation and relaxed geometry}
We first confirmed that all QE results are compatible with VASP results in terms of the electronic structure. Whilst the difference in pseudopotentials induces some minor shifts of the electronic levels the overall ground states are in good agreement between VASP and QE (see  Table~\ref{Table_charges_QE}). Additionally,  the generic features of the computed density of states are the $d$-orbital well separated from the Fermi level and $s$ density close to the Fermi level with opposing contribution (i.e. depolarizing). The exact $d$-orbital position and splitting differs between QE and VASP calculations most likely due to differences in the Hubbard-U correction (i.e. projectors of QE PAW pseudopotentials not being comparable to the VASP ones), however the DOS results qualitatively agree between the two codes. The values of Table~\ref{Table_charges_QE} are also in good agreements, the main difference coming from the different way of calculating charges and magnetic moments, while in QE the L\"owdin charges are used, VASP uses the atomic ones into the PAW Wigner-Seitz sphere. The largest difference is in the magnetic moment, where the VASP calculations have been used to compute the Ti Bader surface and the corresponding localized magnetic moment.

\begin{table*}[h]
\centering
\begin{tabular}{ c| c| c| c| c }
& Total Charge & 3$d$ & 4$s$ & $M_\mathrm{Ti}$ ($\mu_{\rm B}$) \\ 
 \hline
 Ti$_{\rm B}^{(2)}$ & 3.02 (3.322)  & 2.31 (2.444) & 0.705 (0.503)  & 0.87 (1.31) \\  \hline
 Ti$_{\rm O}^{(2)}$ & 3.15  (2.945) & 2.48 (2.270) & 0.71 (0.497) & 2.32 (2.28) \\  \hline
 Ti$_{\rm B}^{(3)}$ & 3.04  (3.240)  & 2.3  (2.388)& 0.73 (0.507) & 1.25 (1.09) \\ \hline 
 Ti$_{\rm O}^{(3)}$ & 3.16  (2.922) & 2.45  (2.235) & 0.71 (0.526) & 2.32 (2.10) \\
\end{tabular}
\caption{\textbf{Orbital occupations of Ti on MgO/Ag(100):} The table reports the total ($s$+$d$) valence charge from L\"owdin charge analysis, the orbital-resolved occupations of the 3$d$ and 4$s$ orbitals, and the total magnetic moment of Ti adsorbed on different MgO sites and thicknesses. Values in parenthesis are from the VASP calculations computed for the PAW Wigner-Seitz sphere, except for the magnetic moment that the Ti Bader surface is used.}
\label{Table_charges_QE}
\end{table*}

%\begin{figure} 
%\includegraphics[width = \linewidth]{./Figures_SI/%pdos-Ti-QE.png}  
%\caption{\textbf{PDOS plots for Ti on Ag/MgO} shown %are the orbital-resolved $s$ and $d$ projected Ti %DOS and $p$ orbital oxygen contributions for 2 and %3 ML of MgO for both adsorption sites. } 
%\label{SI:QE_PDOS}
%\end{figure}

QE calculations used PAW pseudopotentials from the PSL library 1.0.0 with an energy cutoff for the plane-wave expansion of 70 Ry (approx. 900 eV). Exchange-correlation was treated using GGA-PBE and dispersive forces were included using Grimme-D3. This keeps consistency between QE and VASP. We used the QE calculated bulk lattice constant of silver ($a_{\mathrm{Ag}}=4.09$ Angstrom) to build all supercells and assumed that MgO grows epitaxially. All cells were padded by about 2 nm of vacuum in $z$-direction and a dipole decoupling scheme was used to minimize interaction between adjacent periodic images. We used an effective Hubbard $U-J=2$ eV on the 3$d$ manifold to correct some of the common problems of DFT for localized states. The calculations used $4\times4\times1$ k-point sampling. The multiplet calculations~\cite{Wolf_C_and_Delgado_F_2020} are based on a Wannier basis set. Wannierization was performed on a non-spinpolarized GGA calculation using the same parameters as discussed above on an 11x11x1 k-grid using projections on the $s$ and $d$ manifold of Ti and the $p$ manifold of oxygen atoms. The $d$ orbitals result in highly localized Wannier functions (quadratic spread $\Omega^2\approx 1$~ \AA$^2$), whilst the $s$-orbital is more diffuse ($\Omega^2\approx 3$~\AA). The resulting basis was truncated to include 6 states (one $4s$ and five $3d$) to study the orbital arrangement and the behavior of the system on the $s-d$ splitting.

\subsection{Multiplet calculations}
\begin{figure} 
\includegraphics[width = 8cm]{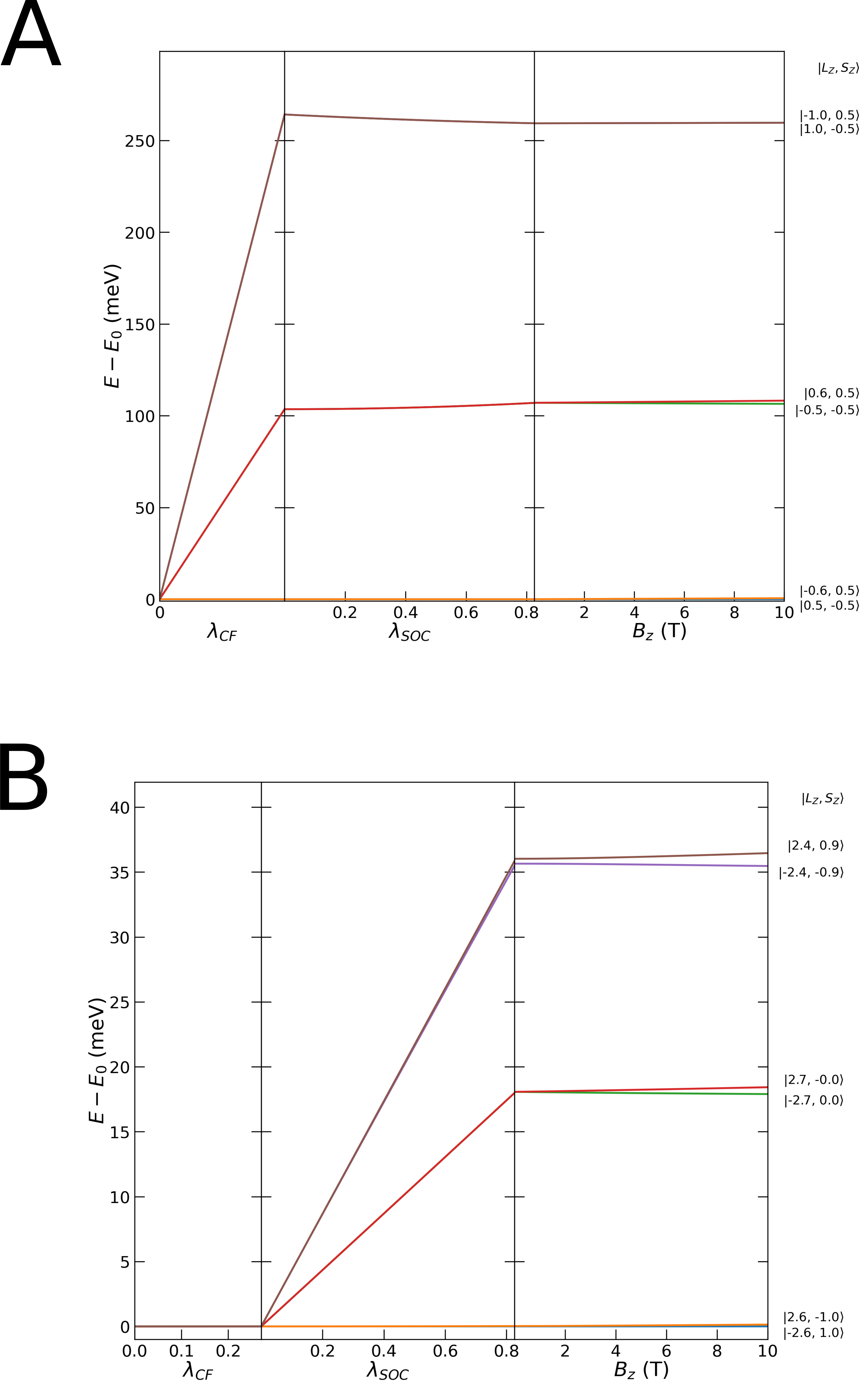}  
\caption{\textbf{Multiplet calculations for Ti on 2 and 3 ML MgO} (A) multiplet calculations for Ti on 2 ML MgO in a 3$d^1$ configuration. (B) the same but with a 3$d^2$ orbital configuration for Ti.} \label{figS5}
\end{figure}

Multiplet calculations~\cite{Wolf_C_and_Delgado_F_2020} were performed based on DFT relaxed geometries for 2 and 3 layers of MgO; The relaxed DFT geometry is used to generate the crystal field positions whilst crystal field (CF) and spin-orbit coupling (SOC) are treated as adjustable parameters using $\lambda_{\mathrm{CF, SOC}}$. The full multiplet spectra including contributions from crystal field, spin-orbit coupling and external magnetic field are shown in Fig.~\ref{figS5}. The results for 2 ML MgO show a ground-state doublet with non-zero orbital moment (responsible for the anisotropy between in and out-of-plane direction). The next doublet is split off by about 100 meV (depending on the exact strength of the crystal field). For 3 ML MgO and a 3$d^2$ orbital configuration the ground state is a $S=1$-like system with strong axial anisotropy due to the underlying oxygen. The degenerate ground-state $|L_Z=\pm3, S_z=\pm1, 0\rangle$ is split into three branches with $|L_Z=\pm3, S_z=\pm1\rangle$ roughly being the ground state (the orbital moment being only slightly quenched) and $|L_Z=\pm3, S_z=0\rangle$ the first excited state split off by spin-orbit coupling.

\clearpage
\newpage

\end{document}